\documentclass[twocolumn,english,prl,showpacs,preprintnumbers]{revtex4-1}
\usepackage{color}
\usepackage{amsmath}
\usepackage{amssymb}
\usepackage{graphicx}
\usepackage{tikz}
\newcommand*\circled[1]{\tikz[baseline=(char.base)]{\node[shape=circle,draw,inner sep=1pt] (char) {#1};}}

\makeatletter

\@ifundefined{textcolor}{}
{%
 \definecolor{BLACK}{gray}{0}
 \definecolor{WHITE}{gray}{1}
 \definecolor{RED}{rgb}{1,0,0}
 \definecolor{GREEN}{rgb}{0,1,0}
 \definecolor{BLUE}{rgb}{0,0,1}
 \definecolor{CYAN}{cmyk}{1,0,0,0}
 \definecolor{MAGENTA}{cmyk}{0,1,0,0}
 \definecolor{YELLOW}{cmyk}{0,0,1,0}
}

\usepackage{textcomp}
\usepackage{dcolumn}
\usepackage{bm}
\usepackage[right]{eurosym}
\usepackage{float}
\usepackage[english]{babel}
\usepackage{blindtext}

\makeatother

\usepackage{babel}
\begin{document}

\title{Interatomic Coulombic decay in helium nanodroplets}

\author{M. Shcherbinin$^{1}$, A. C. LaForge$^{1}$, V. Sharma$^{2}$, M. Devetta$^{3}$, R. Richter$^{4}$, R. Moshammer$^{5}$, T. Pfeifer$^{5}$, M. Mudrich$^{6}$\email{mudrich@phys.au.dk}}

\affiliation{$^{1}$Physikalisches Institut, Universit{\"a}t Freiburg, Germany}

\affiliation{$^{2}$Indian Institute of Technology Hyderabad, India}

\affiliation{$^{3}$Istituto di Fotonica e Nanotecnologie -- Milano, Italy}

\affiliation{$^{4}$Elettra Sincrotrone, Trieste, Italy}

\affiliation{$^{5}$Max-Planck-Institut f{\"u}r Kernphysik, Heidelberg, Germany}

\affiliation{$^{6}$Department of Physics and Astronomy, Aarhus University, Denmark}


\begin{abstract}
Interatomic Coulombic decay (ICD) is induced in helium (He) nanodroplets by photoexciting the $n=2$ excited state of He$^+$ using XUV synchrotron radiation. 
By recording multiple coincidence electron and ion images we find that ICD occurs in various locations at the droplet surface, inside the surface region, or in the droplet interior. ICD at the surface gives rise to energetic He$^+$ ions as previously observed for free He dimers. ICD deeper inside leads to the ejection of slow He$^+$ ions due to Coulomb explosion delayed by elastic collisions with neighboring He atoms, and to the formation of He$_k^+$ complexes.
\end{abstract}


\date{\today}

\maketitle

Isolated atoms or molecules excited by energetic radiation typically decay through intramolecular processes such as the emission of an electron or photon. In contrast, in weakly bound complexes, locally generated electrons can additionally interact with neighboring atoms or molecules, leading to new interatomic or intermolecular interactions. Interatomic Coulombic decay (ICD) is a particularly interesting decay process which occurs when local electronic decay is energetically forbidden~\cite{Cederbaum:1997}. Thus, ICD offers a new, ultrafast decay path where energy is exchanged with a neighboring atom leading to its ionization. Since its discovery, ICD has been observed in a wide variety of weakly-bound systems from He dimers~\cite{Havermeier:2010,Sisourat:2010} and rare-gas clusters to biologically relevant systems such as water clusters; for reviews see~\cite{Hergenhahn:2011,Jahnke:2015}. Today, the focus is on condensed-phase systems where ICD is involved in complex relaxation mechanisms~\cite{Mucke:2010,Nagaya:2016,Stumpf:2016}, which can generate genotoxic low-energy electrons and radical cations~\cite{Boudaiffa:2000}. Recently, it was suggested to utilize this property of ICD for cancer treatment~\cite{Gokhberg:2014,Trinter:2014}.

Here we present the first study of ICD in helium (He) nanodroplets. He nanodroplets are generally considered as an ultracold, inert spectroscopic matrix for embedded, isolated molecules and clusters~\cite{Toennies:2004,Stienkemeier:2006}. Upon ionization by intense or energetic radiation, however, He droplets turn into a highly reactive medium, inducing reactions and secondary ionization processes of the embedded species~\cite{Mudrich:2014}. 
Their homogeneous quantum liquid density profile, and the simple structure of atomic constituents, make He droplets particularly beneficial as benchmark systems for elucidating correlated decay processes. Recent examples include the collective autoionization of multiply excited pure He droplets~\cite{LaForge:2014,Ovcharenko:2014} and the creation of doubly charged species by one-photon ionization of doped He droplets~\cite{LaForgePRL:2016}.  In this work we fully characterize the product states generated by ICD and secondary processes in He nanodroplets using coincidence imaging techniques. 

The experiments were performed using a He droplet machine attached to a velocity map imaging photoelectron-photoion coincidence (VMI-PEPICO) detector at the GasPhase beamline of Elettra-Sincrotrone Trieste, Italy. The apparatus has been described in detail elsewhere~\cite{Buchta:2013,BuchtaJCP:2013}. Briefly, a beam of He nanodroplets is produced by continuously expanding pressurized He (50~bar) of high purity He out of a cold nozzle (10-28~K) with a diameter of 5~$\mu$m into vacuum. At these expansion conditions, the mean droplet sizes range between $\langle N\rangle = 700$ and $\sim 5\times 10^6$ He atoms per droplet. 
In the main detector chamber, the He droplet beam crosses the synchrotron beam perpendicularly in the center of a combined VMI and time-of-flight (TOF) detector. By detecting either electrons or ions with the VMI detector in coincidence with the corresponding particles of opposite charge with the TOF detector, we obtain either ion mass-correlated electron spectra or mass-selected ion kinetic energy (KE) distributions by Abel inversion of the VMIs~\cite{Dick:2014}. 
The XUV photon energy is tuned near the first excited level of He$^+$, $h\nu\gtrsim E(\mathrm{He}^{+\ast},\, n=2)=65.4$~eV~\cite{Havermeier:2010}. 

\begin{figure}
	\centering \includegraphics[width=0.4\textwidth]{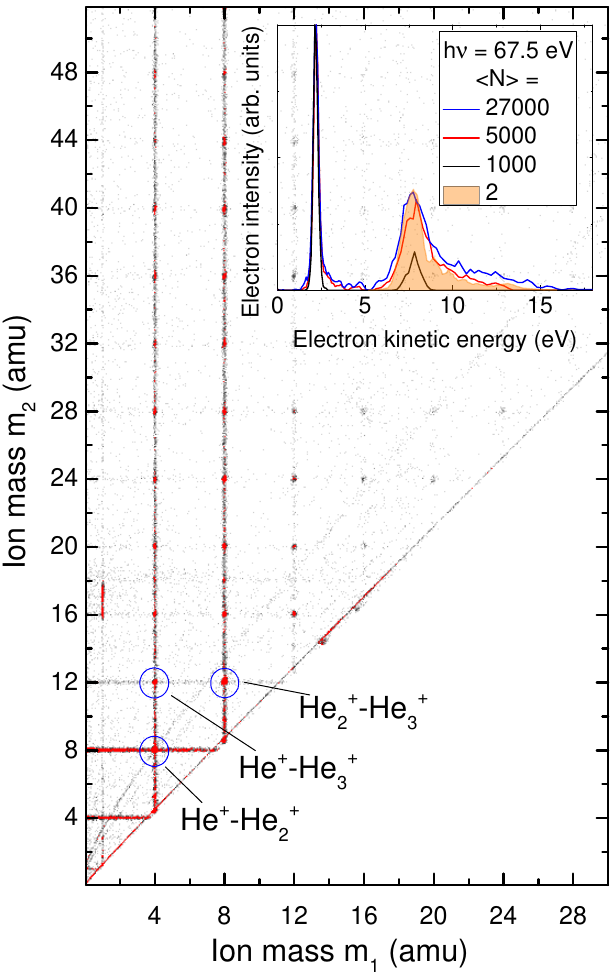}
	\protect\caption{\label{fig:massmap}Ion-ion-electron coincidence time-of-flight mass spectrum recorded at $h\nu = 67.5$~eV and for a mean droplet size of $\langle N\rangle =4000$ He atoms. The inset shows photoelectron spectra measured in coincidence with He$^+$ for various $\langle N\rangle$ and for free He$_2$~\cite{Sisourat:2010}.}
\end{figure}
The elementary ICD process
\[
\mathrm{He}_2 + h\nu \rightarrow \mathrm{HeHe}^{+\ast} + e^-_\mathrm{sat} \rightarrow \mathrm{He}^+ + \mathrm{He}^+ + e^-_\mathrm{sat} + e^-_\mathrm{ICD}
\] 
generates two electrons and two He$^+$ ions flying apart due to Coulomb repulsion. Here, He$^{+\ast}$ denotes a He ion in an excited state with principal quantum number $n>1$. The satellite photoelectron $e^-_\mathrm{sat}$ is emitted directly with kinetic energy $E_\mathrm{sat}=h\nu-E(\mathrm{He}^{+\ast})$ upon simultaneous ionization and excitation of a He atom. The ICD electron $e^-_\mathrm{ICD}$ is created by energy transfer from He$^{+\ast}$ to the neighboring He atom resulting in $E_\mathrm{ICD}=E(\mathrm{He}^{+\ast}) - 2\times E_i - \mathrm{KER}\sim 8.5$~eV~\cite{Havermeier:2010}. Here, $E_i=24.6$~eV denotes the ionization energy of He and $\mathrm{KER}$ is the KE release of the He$^+$-He$^+$ fragments. 

When ICD takes place in He droplets, the primary process is likely to occur between the ionized atom He$^{+\ast}$ and its nearest neighbor due to the steep dependence of the ICD rate on interatomic distance~\cite{Sisourat:2010}. Three-body effects and more complex interactions give  only small contributions~\cite{Sisourat:2016}. However, in He droplets the outgoing ions can interact with the surrounding He atoms and eventually form stable ionic complexes He$^+_k$~\cite{Callicoatt:1996,Shepperson:2011}. 

The simultaneous formation of two He$^+_k$ ions is indeed clearly observed. Fig.~\ref{fig:massmap} displays coincidences of one electron and two ions with masses $m_1$ (horizontal axis) and $m_2$ (vertical axis) as bright spots. The visible lines between integer values are due to false coincidences. While we see He$^+_k$ progressions up to $k=36$ for $\langle N\rangle\sim 30,000$, the most abundant ion-ion coincidences are those of the smallest ions He$^+_{1-3}$, highlighted by circles. Unfortunately, coincidences involving two identical ion masses cannot be resolved with our setup.

Photoelectron spectra recorded in coincidence with ions He$^+$ and He$^+_{2\mathrm{-}3}$ (not shown) at $h\nu = 67.5$~eV for various $\langle N\rangle$ strongly resemble one another and closely match that of free He$_2$, see inset in Fig.~\ref{fig:massmap}. The shown spectrum for free He$_2$ is obtained from the measured KER distribution using the unique relation between KER and photoelectron energies given by the Coulomb potential~\cite{Sisourat:2010}. The sharp line at 2.2~eV represents $e^-_\mathrm{sat}$ and the asymmetrically broadened feature extending from 6 to 16~eV reflects $e^-_\mathrm{ICD}$ created by ICD at various inter-atomic distances~\cite{Sisourat:2010}. The close resemblance of the ICD feature measured in droplets and that of free He$_2$ confirms that ICD proceeds as a binary process with little effect of the droplet on the outgoing electron.
\begin{figure*}
	\centering\includegraphics[width=0.85\textwidth]{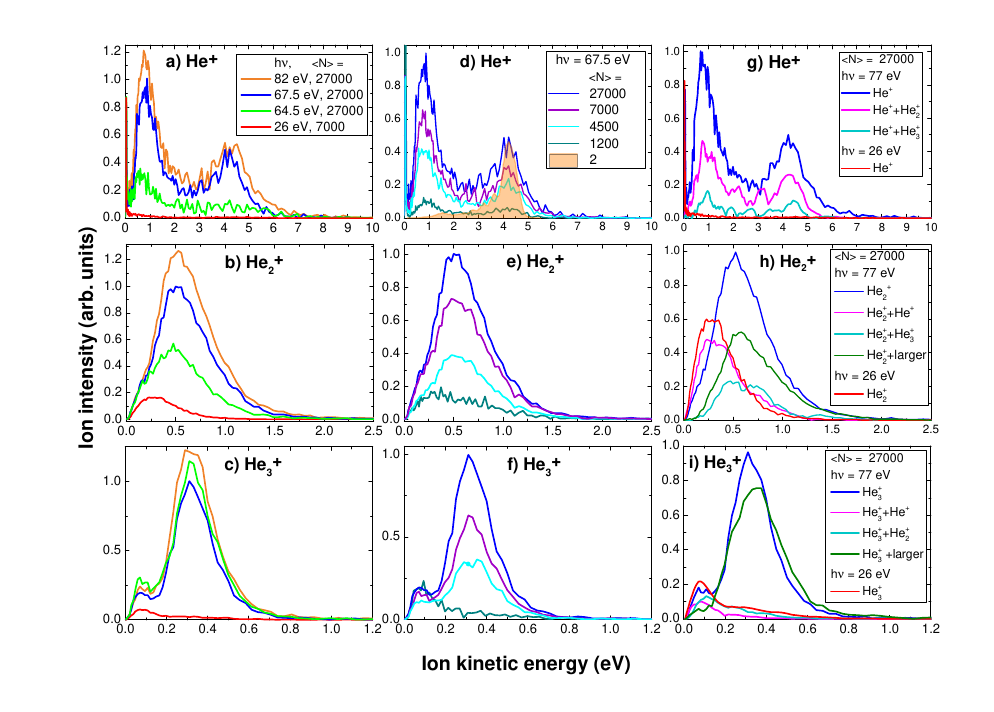}
	\protect\caption{\label{fig:ions}Kinetic energy distributions of He$^+_{1-3}$ ions for various photon energies (left column), He droplet sizes (center), and double-ion coincidences (right column). See text for details.} 
\end{figure*}
The crucial influence of the He droplet on the ICD process is revealed by the KE distributions of ions inferred from ion VMIs. Fig.~\ref{fig:ions} shows the mass-selected ion KE of He$^+_k$ complexes recorded for different experimental paramteters.
For comparison, the ion KE spectrum measured for free He$_2$ at $h\nu= 68.86$~eV is shown in Fig.~\ref{fig:ions} d)~\cite{Havermeier:2010}. The distribution peaked around 4.2~eV is attributed to KER from Coulomb explosion of the pair of He$^+$ ions generated by ICD~\cite{Havermeier:2010,Sisourat:2010}. 
The KE distributions of He$^+$ ions measured with droplets feature a slightly broader structure in the same energy range. 
Thus, part of the He$^+$ ions created by ICD of pairs of He atoms in He droplets are emitted nearly unperturbed. This is most likely to occur at the droplet surface where the He density is low.

\begin{figure}
	\centering \includegraphics[width=0.45\textwidth]{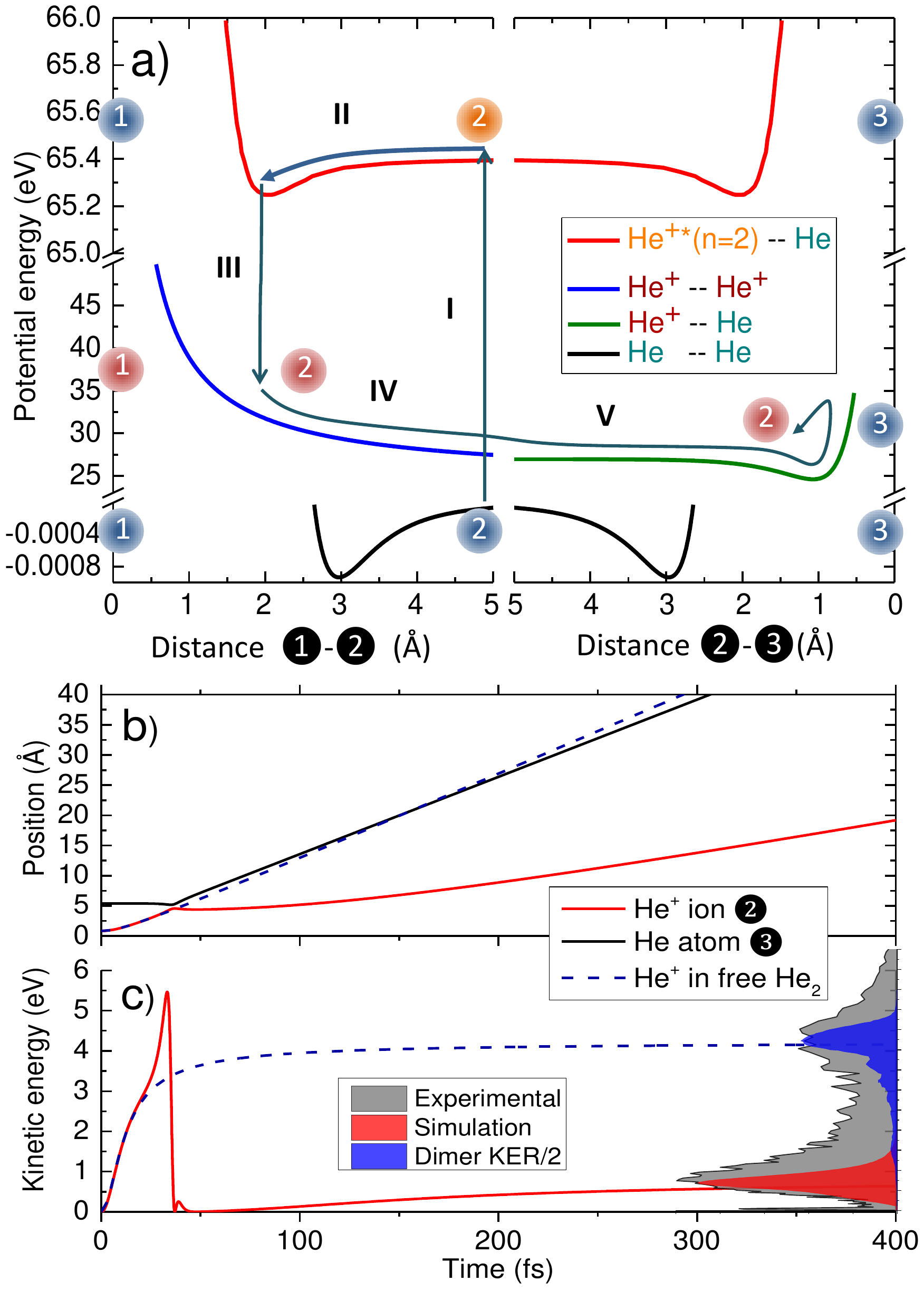}
	\protect\caption{\label{fig:scheme} a) Schematic potential energy level diagram (see text). b) Classical trajectories of a He$^+$ ion colliding with a neighboring He atoms in the course of Coulomb explosion for the linear configuration  
	He-He$^+$-He$^+$-He. c) Corresponding kinetic energies and the simulated and experimental energy distributions (bottom right). } 
\end{figure}
Aside from this clearly ICD-related feature, the He$^+$ KE spectra contain an additional broad peak at about 1~eV and a very narrow peak near 0~eV. The peak near 0~eV is present for all photon energies exceeding $E_i$, see the spectrum recorded at $h\nu=26$~eV shown in Fig.~\ref{fig:ions} a) as a red (lowest) line. Moreover, it is most dominant in the regime of small He droplets where a substantial fraction of free He atoms accompanies the droplet beam. Thus, we attribute this peak to direct photoionization of atomic He. The broad peak around 1~eV is predominantly due to ICD in He droplets. This can be concluded from comparing with the spectrum recorded slightly below the ICD threshold at $h\nu=64.5$~eV, shown in Fig.~\ref{fig:ions} a) as a green (light gray) line. That spectrum contains neither the peak at 4.7~eV nor the high-energy part around 1~eV; only a peak around 0.7~eV and a broad flat feature extending to about 6.5~eV are present. The origin of the broad structure measured at $h\nu = 64.5$~eV is not unambiguously identified at this point. We speculate that secondary processes, in particular inelastic electron-He collisions leading to the production of a second He ion play a role. Note that this structure is absent for $h\nu = 26$~eV where only direct single droplet ionization can occur. 

The prominent feature around 1~eV in the He$^+$ ion spectra evidences  efficient energy loss for He$^+$ ions in droplets, as the coincidence electron spectra show no indications for a corresponding upshift in energy. Obviously,  friction-like multiple elastic scattering of He$^+$ with He atoms inside the droplets may lead to He$^+$ energy loss. However, the ratio of peak integrals of the feature around 1~eV in proportion to that at 4.7~eV only slightly rises from 2.2 to 2.7 when varying the He droplet size $\langle N \rangle$ from 700 to $5\times 10^6$. In contrast, the ratio of the number of He atoms in the bulk of the droplets to those in the surface region ($<90\,\%$ of bulk density) increases from 2 to 54~\cite{Harms:1998}. Thus, the 1~eV feature must be related to ICD occuring in the surface region of the droplets.

What is the origin of the massive loss of KE of He$^+$ when ICD occurrs in He droplets?
We propose the following mechanism, illustrated in Fig.~\ref{fig:scheme} a): Initially one He$^{+\ast}$ ion (labeled \circled{2}) is excited in step I, approaches a neighboring He atom \circled{1} in step II, and decays by ICD (III). In the Coulomb explosion of He$^+$ ions \circled{1} and \circled{2} (IV), each He$^+$ ion flies away from the other until it reaches its neighboring neutral He atom \circled{3} located in the line of flight. There, an energetic billiard-like collision takes place in which the He$^+$ ion transfers its KE to the He atom and thus stops moving if the collision is central (V). Subsequently, Coulomb explosion of the two He$^+$ ICD ions restarts from a larger distance as if ICD occured between non-nearest neighbors~\cite{Fasshauer:2016}, giving rise to a lower final KE. 

This model is supported by a classical trajectory simulation for a linear configuration of atoms He-He$^+$-He$^+$-He. Fig.~\ref{fig:scheme} b) shows the trajectory of He$^+$ ion \circled{2} as a red (lowest) solid line, and of the neighboring He atom \circled{3} as a black (upper) solid line for initial distances between neutral atoms of 3.6~\AA~and between the ICD ions of 1.7~\AA, respectively~\cite{Peterka:2007,Sisourat:2010}. In contrast to freely moving He$^+$ ions (dashed line), in the linear four-atom system, a central collision takes place at $t=37$~fs. The corresponding ion KE, shown in Fig.~\ref{fig:scheme} c), is massively reduced by the collision and converges towards 0.8 eV, in good agreement with the experimental finding. When we run this simulations for a distribution of initial distances between He$^+$ ions given by the measured KER spectrum of the free He$_2$~\cite{Sisourat:2010}, and for a distribution of initial He-He distances corresponding to the He density distribution for $\langle N\rangle =1000$~\cite{Harms:1998}, we obtain the red (lower) smooth spectral feature shown on the right hand side of Fig.~\ref{fig:scheme} c). It nicely matches the low-energy edge of the 1~eV-feature in the experimental droplet spectrum. To simulate the high-energy part, non-central as well as many-body collisions would have to be included, which falls beyond the scope of this work. When determining the initial He-He distance distribution we assume the active surface layer for the described collision process to be constrained towards the bulk of the droplet by the mean free path of He$^+$ in He droplets of 3~\AA, inferred from the gas-phase elastic collision cross section~\cite{Cramer:1957}. Since the He density distribution inside this layer only weakly varies with $\langle N\rangle$, the simulated energy distribution is robust against variations of $\langle N\rangle$.


In case ICD occurs deeper inside the droplets, ICD is followed by He$_k^+$ ion complex formation. 
This we conclude from the sharply rising ratio of detected He$_k^+$ to He$^+$ ions for $k>1$ from 0.4 to 3.4 in the range $\langle N \rangle = 700$ to $5\times 10^6$. Most likely ion complex formation is assisted by elastic stopping collisions to generate slow He$^+$ ions surrounded by He atoms as a precursor. 

The He$^+_2$ and He$^+_3$ KE distributions strikingly differ from those of He$^+$ in that only low energy ions ($\lesssim 2$~eV) are present, see Fig.~\ref{fig:ions} b)-i). This is in line with the concept that He$^+_k$ complexes are formed from the stopped He$^+$ by subsequent aggregation of He atoms inside the droplet. Similar to the low-energy part of the He$^+$ KE spectra [Fig.~\ref{fig:ions} a)], the He$^+_2$ and He$^+_3$ spectra feature two partially overlapping peaks. However, for He$^+_2$ and He$^+_3$ the low-energy component ($\sim 0.3$~eV for He$^+_2$, $\sim 0.1$~eV for He$^+_3$) is already present when singly ionizing the droplets at $h\nu = 26$~eV [red (lowest) lines in Fig.~\ref{fig:ions} b) and c)]. In contrast to atomic He$^+$, He$^+_{k}$ ionic complexes ($k>1$) can be ejected out of neutral He droplets with substantial KE$\lesssim 2.3$~eV released by the stabilization of the complexes in deeply bound vibrational levels~\cite{Buchenau:1991}. The components at higher KE ($\sim 0.6$~eV for He$^+_2$, $\sim 0.35$~eV for He$^+_3$) are already present at $h\nu = 64.5$~eV where ICD is not energetically allowed, but electron impact ionization can create a second ion in the same droplet [green (light grey) lines in Fig.~\ref{fig:ions} b) and c)]. Thus, these parts of the spectra appear to be related to the formation of two He$^+_{2,3}$ ions in the same He droplet, either by ICD or by electron impact ionization. Accordingly, for small droplets with $\langle N\rangle = 1200$ [turquoise (lowest) lines in Fig.~\ref{fig:ions} e) and f)], these components are significantly reduced because electron-impact ionization is improbable and ICD is likely to occur near the droplet surface where at least one ion promptly escapes.

A further confirmation for the 1~eV feature in the He$^+$ KE spectra stemming from ICD is obtained from analyzing the data with regard to multiple ion coincidences. Fig.~\ref{fig:ions} g) shows the KE distributions of He$^+$ detected in coincidence with He$_2^+$ or He$_3^+$ molecular ions [pink and light blue (intermediate) lines], along with He$^+$ single coincidence spectra at $h\nu = 26$ and $77$ eV [blue (upper) and red (lowest) lines]. Aside from differing signal-to-noise ratios, the ion-ion coincidence spectra, which are characteristic for ICD, closely match the single coincidence KE spectrum. 
Thus, even pairs of free ions, He$^++$He$^+$, generated by ICD at the droplet surface,
are subjected to elastic stopping collisions.

In stark contrast to the KE spectra of He$^+$, the He$^+_k$ double ion coincidence spectra for $k=2,\,3$ [Fig.~\ref{fig:ions} i) and h)] clearly differ from one another depending on the size $\ell$ of the second ion detected in the He$^+_k+$He$^+_\ell$ events. While the He$^+_k$ spectra recorded in coincidence with He$^+$ closely match the single droplet ionization spectra at $h\nu = 26$~eV, those recorded in coincidence with larger complexes ($\ell >k-1$) are shifted to higher energies by 0.25-0.35~eV. Consequently, the single-coincidence He$^+_{2,3}$ spectra are superpositions of low and high-energy components, where the high-energy peaks clearly dominate. 
The different energetics of He$^+_k$ ion ejection may arise from the dynamics following ICD. In the case that one He$^+_k$ complex forms inside the droplet and one He$_\ell^+,\,\ell <k$, quickly escapes from it, He$^+_k$ is ejected by vibrational relaxation as in the case of single droplet ionization at $h\nu = 26$~eV. This explains the large low-energy peak in the spectrum of He$^+_3$ recorded in coincidence with He$^+$ and He$^+_2$ [pink and light blue (lowest two) lines in Fig.~\ref{fig:ions} i)]. In contrast, when two complexes He$^+_k +$He$^+_\ell$ form deep inside the same droplet, Coulomb interaction between the two induces more violent dynamics such as mutual repulsion and even droplet fission. 

In summary, we found that in He nanodroplets the primary ICD process occurs as in the free He dimer, and emitted electrons are only weakly perturbed. In contrast, a large fraction of He$^+$ ions undergoes massive energy loss by elastic stopping collisions with neighboring He atoms. Mediated by these collisions, ICD occurring in the droplet interior gives rise to the formation of slow He$_k^+$ complexes, whose energetics crucially depends on the ion escape dynamics. Even more complex reactions might be triggered by such secondary collision processes involving ions and atoms -- not only electrons -- in other condensed-phase systems exposed to energetic radiation.

This work was financially supported by the Deutsche Forschungsgemeinschaft (project MU 2347/10-1). A. C. L. gratefully acknowledges support by the Carl-Zeiss-Stiftung. The authors thank Kirill Gokhberg for stimulating discussions.



%

\end{document}